\documentclass[onecolumn, table, trackchanges]{aastex63}
\graphicspath{{./}{figures/}}

\usepackage[nolist]{acronym}
\usepackage{upgreek}
\usepackage{longtable}

\definecolor{sof}{rgb}{0.4, 0.0, 0.6}

\definecolor{gcb}{rgb}{0.0, 0.6, 1.0}

\definecolor{jrm}{rgb}{0.05, 0.035, 0.45}

\definecolor{mam}{rgb}{0.8, 0.0.4, 0.7}

\newcommand{\seti}{SETI Institute, 339 N Bernardo Ave Suite 200, Mountain View, CA 94043}

\newcommand{\bsrc}{Berkeley SETI Research Center, Berkeley, CA 94720}

\newcommand{\pseti}{Penn State Extraterrestrial Intelligence Center, University Park, PA 16802}

\newcommand{\gbo}{Green Bank Observatory, P.O. Box 2, Green Bank, WV 24944}

\newcommand{\osu}{Department of Astronomy, The Ohio State University, 140 W. 18th Ave., Columbus, OH 43210}

\newcommand{\rit}{School of Physics and Astronomy, Rochester Institute of Technology, Rochester, NY 14623}

\newcommand{\lmarit}{Laboratory for Multiwavelength Astrophysics, Rochester Institute of Technology, Rochester, NY 14623}

\newcommand{\wvu}{Department of Physics and Astronomy and the Center for Gravitational Waves and Cosmology, West Virginia University, Morgantown, WV, 26506-6315}

\newcommand{\giant}{Giant Army, 17th Ave, Seattle, WA 98122}

\newcommand{\gmu}{George Mason University, 4400 University Dr., Fairfax, VA 22030}

\newcommand{\psc}{Penn State Pulsar Search Collaboratory, Department of Astronomy \& Astrophysics, The Pennsylvania State University, 525 Davey Lab, University Park, PA 16802}

\newcommand{\geopsu}{Department of Astronomy and Astrophysics, Department of Geosciences, The Pennsylvania State University, 201 Old Main, University Park, PA 16802}

\newcommand{\uom}{Department of Physics and Astronomy, The University of Maine, Orono, ME 04469}

\newcommand{\umich}{Department of Climate and Space Sciences and Engineering, University of Michigan, Ann Arbor, MI 48109}

\newcommand{\arl}{The Applied Research Laboratory, The Pennsylvania State University, State College, PA, 16802}

\turnoffedit

\begin{document}

\shorttitle{Pulsar Scintillation Bandwidths}
\shortauthors{Penn State PSC}

\title{Scintillation Bandwidth Measurements from 23 Pulsars from the AO327 Survey}

\author[0000-0001-7057-4999]{Sofia Sheikh}
\affiliation{\seti}
\affiliation{\bsrc}
\affiliation{\pseti}

\author[0000-0002-0069-2778]{Grayce C. Brown}
\affiliation{\seti}
\affiliation{\bsrc}

\author[0000-0002-4753-4638]{Jackson MacTaggart}
\affiliation{\umich}

\author[0000-0002-4949-9821]{Thomas Nguyen}
\affiliation{\arl}
\affiliation{\psc}

\author{William D. Fletcher}
\affiliation{\psc}

\author{Brenda L. Jones}
\affiliation{\uom}

\author{Emma Koller}
\affiliation{\psc}

\author{Veronica Petrus}
\affiliation{\psc}

\author{Katie F. Pighini}
\affiliation{\psc}

\author{Gray Rosario}
\affiliation{\psc}

\author[0009-0008-8819-0481]{Vincent A. Smedile}
\affiliation{\osu}

\author{Adam T. Stone}
\affiliation{\geopsu}

\author{Shawn You}
\affiliation{\psc}

\author[0000-0001-7697-7422]{Maura A. McLaughlin}
\affiliation{\wvu}

\author[0000-0002-2451-7288]{Jacob E. Turner}
\affiliation{\gbo}

\author[0000-0003-1226-0793]{Julia S. Deneva}
\affiliation{\gmu}

\author[0000-0003-0721-651X]{Michael T. Lam}
\affiliation{\seti}
\affiliation{\rit}
\affiliation{\lmarit}

\author[0000-0002-7283-1124]{Brent J. Shapiro-Albert}
\affiliation{\wvu}
\affiliation{\giant}

\begin{abstract}

A pulsar's scintillation bandwidth is inversely proportional to the scattering delay, making accurate measurements of scintillation bandwidth critical to characterize unmitigated delays in efforts to measure low-frequency gravitational waves with pulsar timing arrays. In this pilot work, we searched for a subset of known pulsars within $\sim$97\% of the data taken with the PUPPI instrument for the AO327 survey with the Arecibo telescope, attempting to measure the scintillation bandwidths in the dataset by fitting to the 2D autocorrelation function of their dynamic spectra. We successfully measured 38 bandwidths from 23 pulsars (six without prior literature values), finding that: almost all of the measurements are larger than the predictions from NE2001 and YMW16 (two popular galactic models); NE2001 is more consistent with our measurements than YMW16; Gaussian fits to the bandwidth are more consistent with both electron density models than Lorentzian ones; and for the 17 pulsars with prior literature values, the measurements between various sources often vary by factors of a few. The success of Gaussian fits may be due to the use of Gaussian fits to train models in previous work. The variance of literature values over time could relate to the scaling factor used to compare measurements, but also seems consistent with time-varying interstellar medium parameters. This work can be extended to the rest of AO327 to further investigate these trends, highlighting the continuing importance of large archival datasets for projects beyond their initial conception.

\end{abstract}


\section{Introduction}
\label{sec:intro}

Pulsars are stellar remnants, formed via supernova at the end of the life cycle of a sufficiently massive star. Due to conservation of angular momentum and magnetic flux, pulsars have extremely high rotational speeds (with some periods on the order of a millisecond) \citep{lorimer2008binary} and associated strong magnetic fields \citep{ruderman1972pulsars}. As a result of the electromagnetic field strength, energetic particles in the pulsar magnetospheres are accelerated, creating synchrotron radiation which is forced by the rotating fields to the poles of the pulsars, and emitted in jets \citep{lorimer2008binary}. Depending on the geometry of the system relative to Earth, these jets may periodically emit along Earth's line-of-sight. We can detect these jets across the electromagnetic spectrum, most commonly as remarkably regular bursts of dispersed emission in the broadband radio regime; this periodic emission led to the discovery of pulsars by \citet{Hewish1968}.

Inhomogeneities in the ionized \ac{ISM} can cause scattering in the radiation from a broadband radio source, especially one that is compact in spatial extent such as a pulsar, as the radiation propagates through the \ac{ISM}. Scattering causes the radiation to constructively and destructively interfere, producing periodic variations of varying bandwidth and timescale in the radiation's intensity as seen by some distant observer. This effect is called scintillation,  and the relative motions of the observer, the pulsar, and the components of the \ac{ISM} itself \citep{rickett1990radio} will cause changes in other measured parameters such as dispersion measures and fluxes over time. \Ac{DISS} is produced over small spatial scales and is observable on minute to hour timescales at $\sim$GHz frequencies \citep[e.g., ][]{wu2022lofar}, while \ac{RISS} is caused by larger spatial scales and over weeks to months timescales at similar frequencies \citep[e.g., ][]{rickettlyne1990crab}\edit1{\footnote{\ac{DISS} and \ac{RISS} are two regimes in a continuous parameter space, and some science cases do benefit from breaking down the dichotomy by exploring edge cases, or modeling in a purely refractive way \citep[e.g., ][]{jow2023regimes}}}.

The effects of scintillation can be seen in the dynamic spectrum of an observation of a pulsar at radio wavelengths: interference maxima, or ``scintles,'' will appear as bright peaks in the plot of a pulsar's pulse intensity over frequency and time. The characteristic widths in the scintles' appearance in time and frequency can be parameterized by the ``scintillation timescale'' $\Delta t_{D}$ and ``scintillation bandwidth'' $\Delta \nu_{D}$ by measuring the \ac{HWHM} in the \ac{2D ACF} of the dynamic spectrum.  Density fluctuations and turbulence of the \ac{ISM} along the line-of-sight affect both quantities \citep[e.g., ][]{Wang2005}. Additionally, observations that are long enough to derive both the scintillation bandwidth and timescale from \ac{DISS} can provide the 2D power spectrum to the dynamic spectrum, revealing ``scintillation arcs'' \edit1{\citep{stinebring2001faint, walker2004interpretation, cordes2006theory}} whose properties constrain the distance to a scattering screen and the transverse motion of the pulsar.

The scattering delay $\tau_s$ is the broadening timescale, which causes a mean-shift delay \citep[e.g., ][]{Hemberger2008variability}. $\Delta\nu_D$ relates to $\tau_s$ by:
\begin{equation}
\label{eq:scattering_to_bandwidth}
    2 \pi \tau_s \Delta \nu_D = C
\end{equation}
where $C$ is a factor $\sim$1--2 that depends on the geometry and electron density wavenumber spectrum. If a thin screen is assumed, $C$ = 1 for a medium which follows a square law structure function, and $C$ = 0.96 for a medium which follows a Kolmogorov-type structure function \citep{Cordes1998DISS}.

These observational signatures can be used to derive the properties of the \ac{ISM} on the line-of-sight between Earth and a scintillating pulsar \citep[e.g., ][]{Wang2005}, calculate precise orbits of binary pulsars \citep[e.g., ][]{reardon2020precision}, and, with the relation to scattering delay defined above, generate timing corrections for gravitational-wave characterization using pulsar timing arrays \citep[e.g., ][]{liu2022longterm} such as those observed by NANOGrav \citep{arzoumanian2023nanograv, agazie2023nanograv}\edit1{, the EPTA \citep{Chen2021EPTA}, the PPTA \citep{Reardon2021PPTA}, and the InPTA \citep{Tarafdar2022InPTA}}. 

Understanding the distribution of free electrons in the galaxy via modeling the structure of the ionized \ac{ISM} is critical for Galactic composition work, and for understanding the distance to radio sources such as pulsars and \acp{FRB} \citep{price2021pygedm}. One widely used model, NE2001 \citep{cordes2002NE2001}, combined scintillation bandwidth measurements from many pulsars across the celestial sphere with independent distance constraints from e.g., absorption lines in the local hot \ac{ISM}. More recently, \citet{krishnakumar2015scatter} developed an empirical \ac{DM}--scintillation bandwidth relationship that \citet{yao2017new} combined with almost 200 pulsar distances and dispersion measures, as well as HII regions and other observational calibrators, to create the YMW16 model.

However, by using astrometry to provide independent distance measurements for several dozen pulsars, \citet{deller2019microarcsecond} found that NE2001 and YMW16 still have significant flaws, especially at high Galactic latitudes\footnote{Work is being done on YMW16 to mitigate these flaws, e.g., \citet{ocker2020electron}}: some pulsars have significant errors in their DM-derived distances, and both models tend to underestimate the distance to pulsars. Obtaining scintillation bandwidths, and therefore more constraints on electron density and \ac{ISM} structure, can help improve the next generation of electron density models.

One way to do this is by leveraging existing datasets, such as the archival 327-MHz drift-scan survey (AO327) from the sadly collapsed 305-m Arecibo telescope \citep{deneva2013ao327}. However, when using archival data, one must navigate design choices made for other purposes --- in this case, pulsar discovery. Using drift rate data for scintillation work leads to an obvious drawback: the short amount of time that each pulsar is in the beam is much shorter than the scintillation timescale even for DISS. Therefore, we can only use this dataset to extract $\Delta \nu_D$, but cannot place constraints on $\Delta t_D$. $\Delta \nu_D$, however, is the parameter that can be converted to scattering delay, and can be used to help evaluate electron density fluctuations along different sightlines of the galaxy in comparison with models.

In this paper, we describe a search for known pulsars with measurable scintillation bandwidths in AO327, analyze their archival observations, and provide bandwidths and comparisons to electron density model predictions. In Section \ref{sec:obs_and_data} we describe the AO327 survey and the source selection. In Section \ref{sec:analysis}, we describe the analysis pipeline, including \ac{RFI} removal, downsampling, and model fitting. In Section \ref{sec:results}, we discuss each measurement at a per-pulsar level, including a comparison to predictions from existing models, and a discussion of results at a population level. We conclude in Section \ref{sec:discussion_and_conclusion}.

\section{Observations and Data}
\label{sec:obs_and_data}

\subsection{AO327}
\label{ssec:ao327}

AO327 is a drift-scan pulsar survey which was conducted with the 305-m Arecibo telescope at a central frequency of 327 MHz. Details of the campaign design are provided in \citet{deneva2013ao327}, and recent pulsar detections are covered in \citet{deneva2024ao327}. We summarize important parameters here. When taking data in a ``drift-scan'' mode, the telescope remains fixed on a particular azimuth and altitude while the sky rotates overhead: for Arecibo, this led to an effective integration time of 60 seconds for a celestial object to pass through the primary beam, leading to 60 second ``strips'' of coverage in right ascension at the pointing declination. Beginning in 2010, AO327 took data during downtime or unassigned time on Arecibo \citep{deneva2016new}, gathering thousands of hours of data with the Mock Spectrometer from which nearly a hundred pulsars and \acp{RRAT} have been discovered \citep{deneva2013ao327, deneva2016new, martinez2019discovery}.

The \ac{PUPPI} backend, installed in 2012, allowed for the recording of 69 MHz of dual-linear polarization bandwidth, centered at 327 MHz; this bandwidth was channelized into 2816 frequency channels and sampled at 81.92 $\mu$s \citep{deneva2013ao327}. The receiver temperature for the instrument was 113~K, and the gain was 11~K/Jy. For reference, the Mock spectrometer had 0.336~MHz wide frequency channels; the \ac{PUPPI} backend improved on this with a finer, 0.025~MHz channelization. When taken together, the benefits of the survey --- the incredible sensitivity of the Arecibo telescope; the wide bandwidth, relatively low center frequency, and suitable frequency resolution of the \ac{PUPPI} backend; and its huge volume of on-sky time and archival data --- make it a fantastic instrument for measuring scintillation bandwidths. The archival data from AO327 are saved as PSRFITS files, with each individual PSRFITS file containing about 1 hour of data.

\subsection{Data Selection}
\label{ssec:data_selection}

The data used in this work are a subset of the data collected for AO327. The observations span from MJD 56608--59065, and initially included observations taken with both the \ac{PUPPI} and Mock spectrometers. 

To identify pulsars from which we could measure scintillation in frequency, we took each pointing in the available dataset and defined a generous pointing radius of 15$^\prime$, equivalent to twice the \ac{FWHM} of the beam at 327~MHz. We cross-matched these pointings with a list of 223 pulsars that were identified in or discovered by AO327 at the beginning of this work in 2019.

We took those 223 objects and calculated their expected scintillation bandwidths with NE2001, with error bars corresponding to bandwidth calculated at $\pm$20\% of the fiducial distance\footnote{\texttt{pygedm} \citep{price2021pygedm} was used as the NE2001 interface}\edit1{; 20\% represents the distance uncertainty given in \citet{cordes2003ne2001} from, e.g., unmodeled clumps in the \ac{ISM}}. We determined that the narrowest scintles we could resolve with either of the spectrometers on the 327~MHz receiver would be approximately five times their frequency resolution $\Delta \nu$; similarly, we estimated that the \textit{largest} scintle size that we could resolve is approximately half of the bandwidth. We then compared the NE2001 estimates to these values --- if any value within the error bars of the NE2001 estimate fell within the measurable scintle bounds described above, we deemed the pulsar potentially measurable. This filtered our sample to 60 objects. However, 108 objects were additionally, incorrectly, added to the sample due to an error in the input center frequency when calculating the predicted scintillation bandwidths. These 108 objects had true predicted bandwidths that were likely narrower than what we could measure with the AO327 archival data. However, we took them through the same pipeline anyway, giving us 168 pulsars total.

The majority of these pulsars, 128 of the 168, were observed with the more modern PUPPI spectrometer on Arecibo. The Mock spectrometer was used to take a significant fraction of the early AO327 data, but given its limited presence in our sample and its less-optimal parameters for scintillation measurement (i.e., wider channels), we decided to only use PUPPI data in this work for consistency. We searched through $\sim$97\% of the total PUPPI data taken for AO327, with the remaining 3\% representing data taken in the few months before Arecibo's collapse.

\begin{figure}
    \centering
    \includegraphics[width=0.8\textwidth]{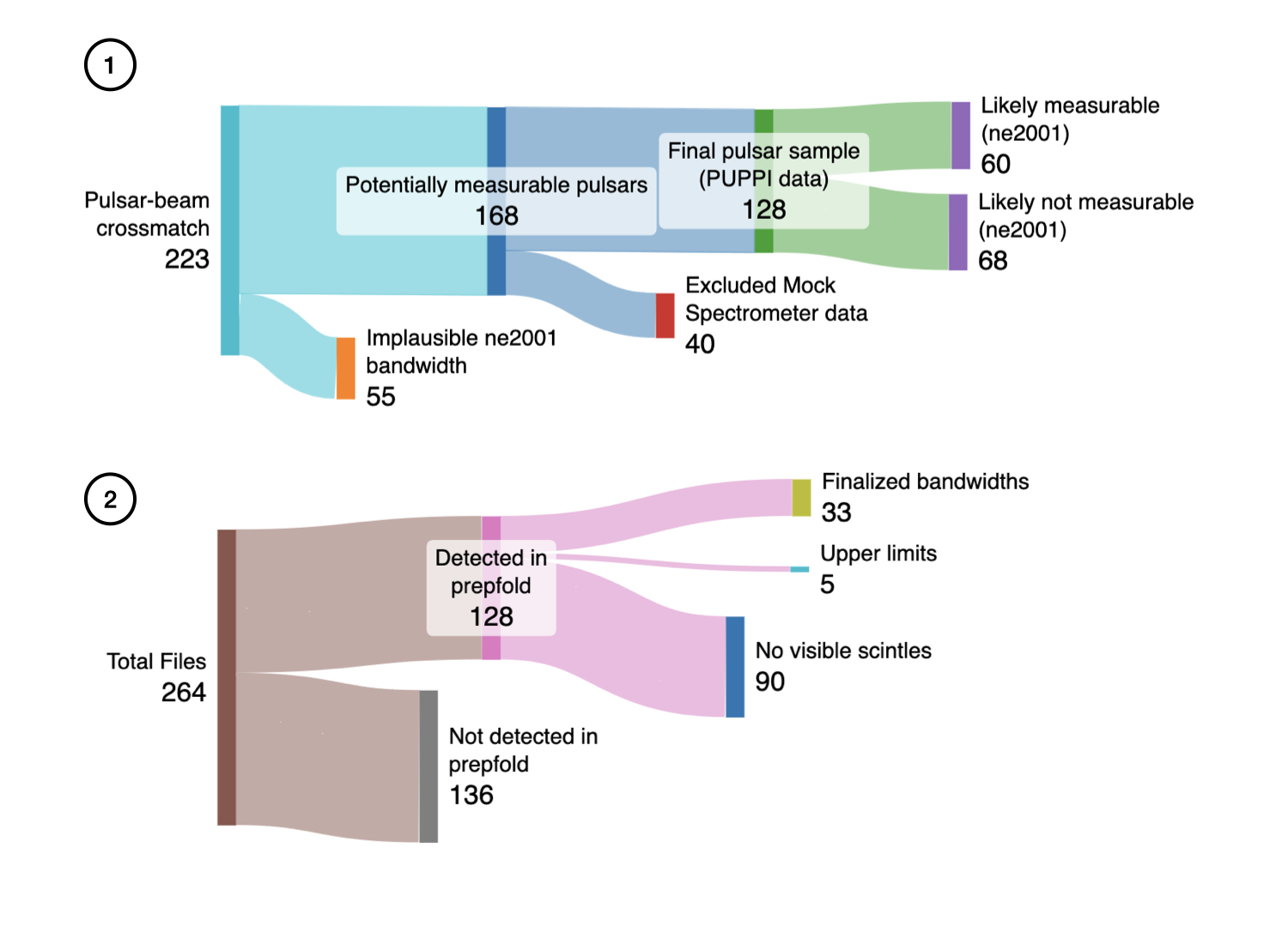}
    \caption{Two sankey diagrams generated with Sankeymatic \citep{sankeymatic} illustrating 1) The path of pulsars from initial identification in the dataset to selection for analysis in Section~\ref{sec:analysis} and 2) The result of each \ac{PUPPI} PSRFITS file from each of those pulsars, leading to the 38 measurements we report in Section~\ref{sec:results}. In subfigure 1), there are two separate NE2001 filtering steps, indicating the original incomplete filtering described in Section~\ref{ssec:data_selection}.}
    \label{fig:file_and_pulsar_path}
\end{figure}

Some of these pulsars had multiple observations in different strips, observed on different days. We include 264 individual 4-bit PSRFITS files, corresponding to scans within a 15$^\prime$ radius of the 128 sample pulsars. Overall, this sample has relatively uniform coverage in right ascension and declinations visible to Arecibo, which is ideal for comparison with electron density models. The full selection path for both pulsars and files is shown in Figure~\ref{fig:file_and_pulsar_path}.

For comparison, recent work in \citet{deneva2024ao327} found 206 pulsars within the AO327 PUPPI data, using a pointing radius of 7.5$^\prime$ (1/2 of what we assumed above). 40\% of pulsars in the \citet{deneva2024ao327} list are included in this work; our sample was determined before the full PUPPI dataset had been searched for pulsar emission, leading to the discrepancy.

Interestingly, this pilot study serendipitously gives us insight into how the different choices of beamwidth affect the recovered pulsar sample. Though the recovery fraction (number of predicted pulsars detected in folded data, as described in Section~\ref{ssec:prepfold}) is lower for our pulsars that are outside the beam's FWHM, we still manage to detect an additional 27 pulsars that would not be considered if using the 7.5$^\prime$ FWHM as the pointing boundary. In future work with the AO327 sample, we recommend cross-matching pulsar catalogs with this larger radius if vetting time allows, in order to recover and characterize pulsars that are entering the beam through the sidelobes\footnote{Arcing \ac{RFI} from geostationary satellites has been observed in Arecibo data, even when the satellites were in the extended sidelobes a few degrees away from the main beam \citep{ferguson2022spectral}}.

\section{Data Analysis}
\label{sec:analysis}

\subsection{Identifying detections with \texttt{prepfold}}
\label{ssec:prepfold}

To identify whether a pulsar was detected in an observation, we folded the data using the \texttt{prepfold} function from \ac{PRESTO}, a large suite of pulsar search programs and software \citep{Ransom2011presto}.  The \texttt{prepfold} program dedisperses and folds pulsar data to create output summary plots within which true pulsar detections are quite obvious compared to \ac{RFI}. We folded our data according to parameters from the ATNF\footnote{\url{http://www.atnf.csiro.au/research/pulsar/psrcat}}\ Pulsar Catalogue \citep{manchester2005atnf}, specifically, period and \ac{DM}. Since we knew each of these particular pulsars should be in the data based on the filtering from Section \ref{ssec:data_selection}, we inputted the known parameters into \texttt{prepfold} instead of having the function search through a grid of \acp{DM} and periods. This method does search a small part of parameter space around the input period and \ac{DM}, allowing for accurate folding even without a full timing ephemeris. It should be noted that this method will not always work for binary pulsars, which constituted 10 of the original 223 objects; the only one of these which had visible scintles was B0820+02, which is discussed in Section \ref{ssec:B0820+02}.

The time-phase plot of the \texttt{prepfold} summary plot is a good diagnostic of whether or not a pulsar was detected. If there is a pulsar in the data, it will appear in the time-phase sub-plot as two dark, vertical lines. Since the length of the observation was much greater than the amount of time the pulsar spent in the beam, we cropped the observation to the times where the pulsar was visibly in the beam. After creating and inspecting the 264 \texttt{prepfold} plots, we removed 136 PSRFITS files from our sample where the pulsar was not detected, leaving 128 files with sufficient pulsar signal to analyze in the next step. The pulsars that we did not detect were generally the ones furthest from the beam (between 1--2 times the \ac{FWHM}) and the intermittent pulsars (i.e., pulsars which do not consistently emit, such as \acp{RRAT} or nulling pulsars), as expected. 

\subsection{Folding}
\label{ssec:folding}

Once we identified the time range containing a pulsar, we created folded output files from the PSRFITS files described in Section \ref{ssec:ao327} using the \texttt{fold\_psrfits} function from \texttt{psrfits\_utils}, a utility library for working with PSRFITS pulsar files\footnote{written by P.~Demorest and S.~Ransom and available at \url{https://github.com/demorest/psrfits_utils}}. We used the same ATNF values from the \texttt{prepfold} commands above to create three output folded PSRFITS files for each pulsar observation, with time integrations of 5, 10, and 20 seconds. Generally, the 20 second integrations had the highest \ac{SNR} and were used in the following steps, but for some observations it was advantageous to use the shorter integrations to mitigate \ac{RFI} and/or provide sufficient time bins for \texttt{PyPulse} to successfully produce a dynamic spectrum (see Section \ref{ssec:pypulse}). Each output file was also reduced to 64 phase bins to increase \ac{SNR} at this step.

\subsection{\texttt{pazi} and fscrunch}
\label{ssec:pazi_and_fscrunch}

We used the \texttt{pazi} tool from \texttt{PSRCHIVE} \citep{hotan2004psrchive} to remove \ac{RFI} from the folded data by zapping frequency channels and time bins with significant interference. The most common RFI signal we observed was a bright, narrowband emitter near 312~MHz which, if not excised, would entirely override the \texttt{PyPulse} pulse identification within the dynamic spectrum in a later step. After using \texttt{pazi}, we reduced the number of frequency bins by ``fscrunching'', or summing adjacent frequency bins to increase \ac{SNR} using the \texttt{pam} utility. In this process, we were cautious not to overscrunch the data beyond either the limit of five datapoints across from NE2001 in Section \ref{ssec:data_selection}, nor to overscrunch empirically based on visible scintles in the dynamic spectrum.

\subsection{\texttt{PyPulse}}
\label{ssec:pypulse}

We used the Python library \texttt{PyPulse} \citep{lam2017pypulse} to create the dynamic spectra which were the basis of the scintillation bandwidth measurements. We initiated each cropped, \ac{RFI}-cleaned, and frequency scrunched file into the \texttt{Archive} class within \texttt{PyPulse}. We also generated a pulse template, created by averaging the file in time, frequency, and polarization, and then smoothing the new profiles with the \texttt{psrsmooth} function of \texttt{PSRCHIVE}. \texttt{PyPulse} uses the data and the pulse template to generate a pulse-sensitive dynamic spectrum which integrates all intensity across the pulse for each ``pixel'' of the dynamic spectrum's time-frequency grid. At this point, if there were any additional time bins without pulsar signal (e.g., fading on the edges due to the pulsar entering or leaving the beam), they were cropped out of the dynamic spectrum to improve \ac{SNR} for the following steps. 

\subsection{2D ACF}
\label{ssec:2DACF}

We then performed a \ac{2D ACF} on the dynamic spectra using the \texttt{acf2d} function in the \texttt{PyPulse} package. Generally, when the ACF method has been used to measure scintillation bandwidth in prior studies \citep[e.g., ][]{Wang2005}, both axes (time and frequency) of the data are lagged and then used for measurement of $\Delta \nu_D$ and $\Delta t_D$. As mentioned in Section \ref{sec:obs_and_data}, our short observation times do not provide any useful measurements across the time axis of the \ac{2D ACF}. Therefore, we take the sum across the time lag axis to produce a ``slice'' of scaled flux density versus frequency lag, which will have a morphology with a central peak of some measurable width. \edit1{We opt to use the sum across the time lag axis in this work because we are dealing with short observation times; the sum across the time lag axis recoups some signal-to-noise and does not suffer here from additional artifacts that would occur with longer observation times. We do acknowledge that this method is not without its own minor drawbacks (e.g., including noise in the outskirts, summing features of differing widths), but do not find them significant enough to affect the results of the work.}

\subsection{Fitting for scintillation bandwidth}
\label{ssec:fitting}

Even after these multiple phases of data reduction, there are a few frequency-dependent effects that may still be present in the \ac{2D ACF}, e.g., rolloff of intensity at the edges of the bandpass. To ensure we are measuring purely pulsar scintillation, we crop the data in the frequency lag axis to contain only the smallest coherent structure in the peak with at least five data points, so as to avoid fitting to only the direct current (DC) spike at lag = 0. \edit1{In cases where the peak is narrower than 5 points across the ACF slice ($<$0.084 MHz), we do not fit a model, and instead report an upper limit, indicating that the actual bandwidth is narrower than we can measure.} We fit this structure with both a Gaussian model (used more commonly in the literature) and a Lorentzian model (a better mathematical fit to the assumed \ac{ISM} structure) using the \texttt{curve\_fit} function from \texttt{scipy.optimize} \citep{2020SciPy-NMeth}.

The Gaussian model is described by:
\begin{equation}
    \label{eq:gaussian}
    A \exp{\bigg(\frac{-(x-\mu)^2}{2\sigma^2}\bigg)}+y
\end{equation}

\noindent
where $A$ determines the amplitude of the Gaussian, $\mu$ determines its location along the frequency-lag axis, $\sigma$ determines its width, and $y$ determines its vertical offset from 0. We set $\mu$ to 0 because the ACF is always guaranteed to be symmetric.

The Lorentzian model is described by:
\begin{equation}
    \label{eq:lorentzian}
    A\left(\frac{w}{x^2 + w^2}\right) + y
\end{equation}

\noindent
where $A$ determines the amplitude of the Lorentzian, $w$ determines its width, and $y$ determines its vertical offset from 0.

From the width parameters, we can convert to a $\Delta \nu_D$ by taking the half-width at half-maximum --- equivalent to the width parameter for the Lorentzian and equal to $\sqrt{2 \ln{2}}$ times the width parameter (standard deviation) for the Gaussian.

\subsection{Calculating errors in the fit}
\label{ssec:errors}

We incorporated three sources of error, added in quadrature, which we report alongside the final scintillation bandwidth measurements\edit1{: \ac{FSE}, fit error, and channel width uncertainty.}

\Ac{FSE} accounts for the fact that there can only be a finite amount of scintles observed in a limited bandwidth. This effect is stronger when the scintles are proportionally larger in the bandpass, due to there being fewer scintles. We calculate \ac{FSE} using \edit1{the method from \citet{cordes1986space}:}

\begin{equation}
   {\rm FSE}_{\Delta \nu,\rm Gauss} = \frac{\Delta \nu_D}{2\ln(2)N_{\rm scint}}
    \label{eq:fse}
\end{equation}

\noindent
\edit1{where $N_{\rm scint}$ is the number of scintles in the observation. $N_{\rm scint}$ is normally defined as the number of scintles in frequency multiplied by the number of scintles in time --- $N_{\nu} N_{\tau}$ --- but here our observation time is much shorter than $\Delta \nu_{D}$ and thus $N_{\tau} = 1$. Thus, using the formulation from \citet{cordes2010measurement}:}

\begin{equation}
    N_{\rm scint} = N_{\nu} = 1 + \frac{\eta_{\nu}B}{\Delta \nu_D}
    \label{eq:fse_nscint}
\end{equation}

\noindent
\edit1{where $\eta_{\nu}$ is a filling factor that we set to 0.2 \citep[as in ][]{cordes2010measurement, turner2021nanograv} and $B$ is the bandwidth of the observation. Equation \ref{eq:fse_nscint} can then be inserted into Equation \ref{eq:fse} to obtain the \ac{FSE}.}

We also included the fit error on the width parameters ($\sqrt{2 \ln{2}}$ times the \edit2{square root of the variance} $\sigma$ for the Gaussian fit, the \edit2{square root of the variance} $w$ directly for the Lorentzian fit) from \texttt{curve\_fit}. To account for the uncertainty in particularly narrow scintles, we also added an additional error term equal to the width of half of a frequency channel after frequency scrunching.

\begin{figure}
    \centering
    \includegraphics[width=\textwidth]{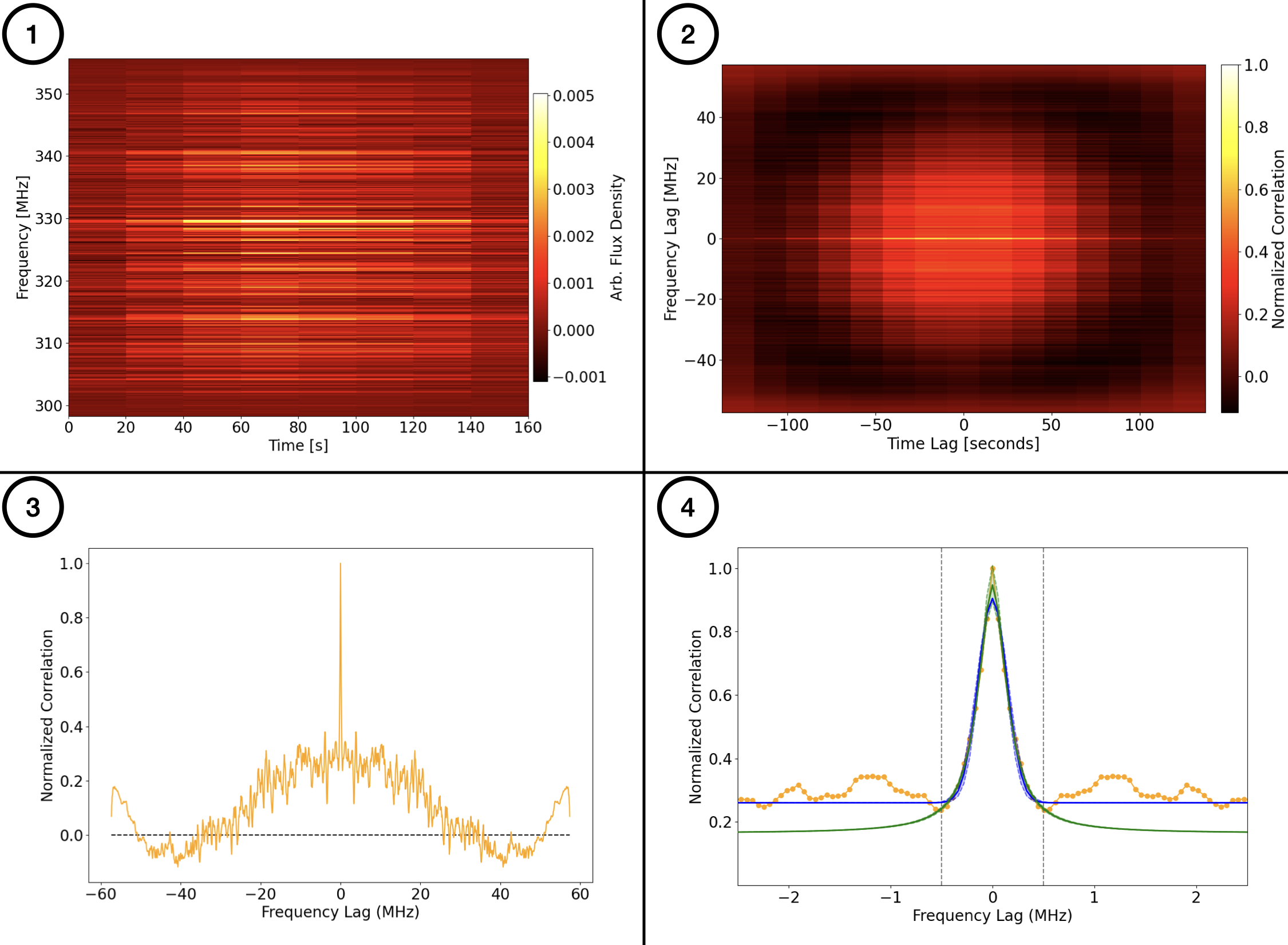}
    \caption{An example of the analysis pipeline described throughout Section~\ref{sec:analysis} for a scan of pulsar B2315+21. \textbf{Subplot 1}: The dynamic spectrum after loading into \texttt{PyPulse} with a pulse template. Time is shown on the horizontal axis and frequency on the vertical axis. The pulsar fades in as it enters the drift scan beam and then fades out as it leaves the beam a few minutes later. Scintillation is visible in the horizontal striping across the pulsar signal from top-to-bottom. \textbf{Subplot 2}: The 2D autocorrelation function (2D ACF) of the dynamic spectrum, showing the correlation of the signal with a delayed copy of itself as a function of lag (delay) in time and frequency. The correlation coefficient is a measure of similarity between the signal and its copy\edit2{, with the central peak normalized to unity}. The frequency lag is shown on the vertical axis and the time lag is shown on the horizontal axis. The horizontal extent (in time lag) only indicates the time the pulsar was in the beam, not the scintillation timescale, and is thus unimportant for this study. The width of the narrowest horizontal stripe in the \ac{2D ACF} is indicative of the scintillation bandwidth corresponding to the stripes in Subplot 1. \textbf{Subplot 3}: The \ac{2D ACF} ``slice,'' created by summing the \ac{2D ACF} in Subplot 2 along the time axis. The narrowest peak in this image can be fit with a Gaussian or Lorentzian to measure the scintillation bandwidth. For some faint pulsars, the bounds of the fit will be determined by crossing points where the correlation coefficient becomes negative (dashed line). \textbf{Subplot 4}: The \ac{2D ACF} slice from Subplot 3, but zoomed in to the middle few MHz and overplotted with both a best-fit Lorentzian (green) and a best-fit Gaussian (blue) \edit2{to the points between the vertical gray dashed lines, using a least squares fitting method}. These are the measurements reported in Table~\ref{tab:measurements}.}.
    \label{fig:analysis_pipeline}
\end{figure}

To visualize the entire analysis process in Section~\ref{sec:analysis}, we summarize four major steps using the output plots from our analysis pipeline in Figure~\ref{fig:analysis_pipeline}.

\section{Results}
\label{sec:results}

Using the analysis procedure from Section \ref{sec:analysis}, we measured 38 scintillation bandwidths, $\Delta \nu_D$, from 23 unique pulsars. In Table \ref{tab:measurements}, we report the $\Delta \nu_D$ values for both Gaussian and Lorentzian fits with their associated error (reported with $\delta$), compare these values to predictions from NE2001 and YMW16\footnote{Note that YMW16 does not directly make predictions for scintillation bandwidth based on structures in the ISM, but uses the empirical relationship derived by \citet{krishnakumar2015scatter}, which is what is used in this work.}, and provide the same information converted to scattering time, $\tau_s$, using the relation defined in Equation \ref{eq:scattering_to_bandwidth}.

\startlongtable
\begin{deluxetable}{lccccccccc}
\tablecaption{The 38 scintillation bandwidth measurements $\Delta \nu_D$ reported in this study, and their equivalent scattering timescales $\tau_{s}$.}
\tablehead{
\colhead{Pulsar} & \colhead{MJD}  & \colhead{$\Delta \nu_{D,\rm Lor}$} & \colhead{$\Delta \nu_{D,\rm Gauss}$} & \colhead{$\Delta \nu_{D,\rm NE2001}$} & \colhead{$\Delta \nu_{D,\rm YMW16}$} & \colhead{$\tau_{s,\rm Lor}$} & \colhead{$\tau_{s,\rm Gauss}$} & \colhead{$\tau_{s,\rm NE2001}$} & \colhead{$\tau_{s,\rm YMW16}$} \\
\colhead{} & \colhead{}  & \colhead{(MHz)} & \colhead{(MHz)} & \colhead{(MHz)} & \colhead{(MHz)} & \colhead{($\upmu$s)} & \colhead{($\upmu$s)} & \colhead{($\upmu$s)} & \colhead{($\upmu$s)}}
\startdata
B0301+19   & 55543 & \textless{}0.084 & \textless{}0.084 & 0.054  & 0.070  & \textgreater{}1.82 & \textgreater{}1.82 & 2.9    & 2.3   \\
B0301+19   & 58237 & \textless{}0.084 & \textless{}0.084 & 0.054  & 0.070  & \textgreater{}1.82 & \textgreater{}1.82 & 2.9    & 2.3   \\
B0820+02   & 58225 & 0.09(1)          & 0.10(1)          & 0.055  & 0.020  & 1.6(2)             & 1.4(2)             & 2.9    & 8.0   \\
B0820+02   & 58977 & 0.07(1)          & 0.07(1)          & 0.055  & 0.020  & 2.1(4)             & 2.1(4)             & 2.9    & 8.0   \\
B0823+26   & 57029 & 0.21(2)          & 0.21(2)          & 0.077  & 0.037  & 0.73(8)            & 0.74(8)            & 2.1    & 4.3   \\
B0834+06   & 58980 & 0.4(1)           & 0.4(1)           & 0.21   & 0.12   & 0.4(1)             & 0.4(1)             & 0.76   & 1.3   \\
B0919+06   & 58915 & 0.15(2)          & 0.15(2)          & 0.018  & 0.013  & 1.0(1)             & 1.0(1)             & 8.8    & 13    \\
B0919+06   & 58923 & 0.13(2)          & 0.10(1)          & 0.018  & 0.013  & 1.2(2)             & 1.5(2)             & 8.8    & 13    \\
B0919+06   & 58980 & 0.12(1)          & 0.13(2)          & 0.018  & 0.013  & 1.2(2)             & 1.2(1)             & 8.8    & 13    \\
B0950+08   & 58848 & 31(18)           & 20(11)           & 101    & 4.0    & 0.005(3)           & 0.007(4)           & 0.0016 & 0.040 \\
B1237+25   & 58287 & 10(5)            & 9(4)             & 0.21   & 0.29   & 0.015(7)           & 0.018(9)           & 0.77   & 0.56  \\
B1237+25   & 55556 & 4(2)             & 4(1)             & 0.21   & 0.29   & 0.04(1)            & 0.04(1)            & 0.77   & 0.56  \\
B1530+27   & 58234 & 0.49(6)          & 0.40(5)          & 0.34   & 0.085  & 0.31(4)            & 0.38(5)            & 0.47   & 1.9   \\
B1929+10   & 58496 & 0.61(9)          & 0.56(8)          & 5.5    & 3.4    & 0.25(4)            & 0.27(4)            & 0.029  & 0.047 \\
B1929+10   & 58673 & 2.0(5)           & 2.0(5)           & 5.5    & 3.4    & 0.07(2)            & 0.08(2)            & 0.029  & 0.047 \\
B1929+10   & 58726 & 1.3(2)           & 1.0(2)           & 5.5    & 3.4    & 0.12(2)            & 0.15(3)            & 0.029  & 0.047 \\
B1952+29   & 57966 & 0.8(4)           & 0.6(2)           & 0.21   & 0.42   & 0.2(1)             & 0.28(8)            & 0.75   & 0.38  \\
B2016+28   & 58055 & 0.13(2)          & 0.11(1)          & 0.038  & 0.093  & 1.1(4)             & 1.4(2)             & 4.2    & 1.7   \\
B2016+28   & 58058 & 0.14(2)          & 0.15(2)          & 0.038  & 0.093  & 1.1(1)             & 1.0(1)             & 4.2    & 1.7   \\
B2020+28   & 58055 & 0.27(3)          & 0.22(2)          & 0.056  & 0.018  & 0.57(6)            & 0.68(7)            & 2.9    & 9.0   \\
B2020+28   & 58056 & 0.28(6)          & 0.27(6)          & 0.056  & 0.018  & 0.5(1)             & 0.6(1)             & 2.9    & 9.0   \\
B2020+28   & 58057 & 0.26(3)          & 0.26(3)          & 0.056  & 0.018  & 0.59(7)            & 0.58(6)            & 2.9    & 9.0   \\
B2110+27   & 56608 & 0.05(1)          & 0.05(1)          & 0.066  & 0.017  & 3.0(7)             & 2.8(7)             & 2.4    & 9.6   \\
B2110+27   & 57004 & 0.07(1)          & 0.07(1)          & 0.066  & 0.017  & 2.1(4)             & 2.3(4)             & 2.4    & 9.6   \\
B2110+27   & 57145 & 0.06(1)          & 0.07(1)          & 0.066  & 0.017  & 2.6(5)             & 2.2(4)             & 2.4    & 9.6   \\
B2110+27   & 58331 & 0.06(1)          & 0.06(1)          & 0.066  & 0.017  & 2.7(6)             & 2.6(6)             & 2.4    & 9.6   \\
B2315+21   & 55546 & 0.17(3)          & 0.17(3)          & 0.090  & 0.030  & 0.9(2)             & 0.9(2)             & 1.8    & 5.3   \\
J0051+0423 & 56795 & 0.47(6)          & 0.41(5)          & 0.078  & 0.099  & 0.32(4)            & 0.38(5)            & 2.1    & 1.6   \\
J0051+0423 & 56868 & 0.35(5)          & 0.33(4)          & 0.078  & 0.099  & 0.44(6)            & 0.47(6)            & 2.1    & 1.6   \\
J0137+1654 & 58423 & 0.03(1)          & 0.02(1)          & 0.018  & 0.014  & 6(3)               & 7(3)               & 8.7    & 12    \\
J1313+0931 & 57037 & 0.35(6)          & 0.31(6)          & 0.088  & 0.15   & 0.44(8)            & 0.49(9)            & 1.8    & 1.1   \\
J1612+2008 & 58251 & 0.32(7)          & 0.25(6)          & 0.029  & 0.037  & 0.5(1)             & 0.6(2)             & 5.4    & 4.3   \\
J1652+2651 & 57159 & \textless{}0.084 & \textless{}0.084 & 0.0053 & 0.0030 & \textgreater{}1.82 & \textgreater{}1.82 & 30     & 53    \\
J1758+3030 & 57753 & \textless{}0.084 & \textless{}0.084 & 0.012  & 0.0053 & \textgreater{}1.82 & \textgreater{}1.82 & 13     & 30    \\
J2215+1538 & 56759 & 0.04(1)          & 0.04(1)          & 0.017  & 0.0099 & 3(1)               & 4(1)               & 9.2    & 16    \\
J2215+1538 & 56848 & \textless{}0.084 & \textless{}0.084 & 0.017  & 0.0099 & \textgreater{}1.82 & \textgreater{}1.82 & 9.2    & 16    \\
J2227+3038 & 58264 & 0.19(2)          & 0.21(2)          & 0.016  & 0.0065 & 0.80(9)            & 0.74(8)            & 9.9    & 24    \\
J2253+1516 & 58638 & 0.03(1)          & 0.03(1)          & 0.016  & 0.010  & 5(2)               & 5(2)               & 10     & 16   \\
\enddata
\label{tab:measurements}
\end{deluxetable}

\label{tab:measurements}

We also quantitatively compare the $\Delta \nu_D$ measurements for each pulsar with their previously-measured values in the literature as summarized in Table \ref{tab:literature}. For each pulsar and each reference in Table \ref{tab:literature}, we provide the equivalent scintillation bandwidth at 327~MHz $\nu_{D, 327}$ (MHz) and the equivalent scattering timescale $\tau_{s,327}$ ($\upmu$s), along with a reference key. For references which originally reported a scattering timescale, we convert from scattering measurements to bandwidth equivalents using Equation \ref{eq:scattering_to_bandwidth}, where we use $C$ = 1.53 to align with \citet{taylor1993catalog} and $C$ = \edit1{0.96 in all other cases to be consistent with a Kolmogorov thin-screen model}. For all measurements, after converting from scattering timescale to bandwidth as necessary, we scale from the original measurement frequency to 327~MHz using a Kolmogorov scaling index of $\alpha$ = 4.4 as in \citep{rickett1977interstellar}. \edit1{It should be noted that scaling indices have been found to vary significantly from 4.4, generally tending smaller \citep[e.g., ][]{levin2016nanograv}, as that value is only true for an ideal screen where inner scale and refractive effects are ignored \citep{turner2021nanograv}.}


\startlongtable
\begin{deluxetable}{|l|lll|l|lll|}
\tablecaption{A summary of all literature values for the 23 pulsars included in this study.}
\tablehead{
\colhead{Pulsar} & \colhead{$\nu_{\rm scint, 327}$} & \colhead{$\tau_{sc,327}$} & \colhead{Ref.} & \colhead{Pulsar} & \colhead{$\nu_{\rm scint, 327}$} & \colhead{$\tau_{sc,327}$} & \colhead{Ref.} \\ 
\colhead{} & \colhead{(MHz, conv.)} & \colhead{($\upmu$s, conv.)} & \colhead{} & \colhead{} & \colhead{(MHz, conv.)} & \colhead{($\upmu$s, conv.)} & \colhead{}
}
\startdata
B0301+19   & 0.0455                            & 3.36              & KLL       & B1237+25   & 0.89             & 0.17              & Z         \\
           & 0.0387                            & 3.94              & TML       &            & 1.5              & 0.10              & S\&W      \\
           & 0.23                              & 0.67              & S\&W      &            & 0.334            & 0.457             & C\&L 2003 \\
           & 0.062                             & 2.5               & C\&L 2003 &            & 6.1              & 0.025             & This Work \\
           & \textless{}0.084                  & \textgreater{}1.8 & This Work & B1530+27   & 0.0223           & 6.85              & KLL       \\
B0820+02   & 0.0387                            & 3.94              & TML       &            & 0.173            & 0.883             & TML       \\
           & 1.5                               & 0.10              & S\&W      &            & 0.197            & 0.776             & BGR       \\
           & 0.062                             & 2.5               & C\&L 2003 &            & 0.63             & 0.24              & Z         \\
           & 0.013                             & 12                & This Work &            & 0.278            & 0.550             & C\&L 2003 \\
B0823+26   & 0.134                             & 1.14              & KLL       &            & 0.40             & 0.38              & This Work \\
           & 0.0435                            & 3.52              & TML       & B1929+10   & 0.628            & 0.243             & TML       \\
           & 0.057                             & 2.7               & DLK       &            & 0.83             & 0.18              & S\&W      \\
           & 0.293                             & 0.521             & BGR       &            & 1.01             & 0.151             & C\&L 2003 \\
           & 0.34                              & 0.45              & Z         &            & 1.2              & 0.13              & This Work \\
           & 0.19                              & 0.81              & S\&W      & B1952+29   & 0.165            & 0.925             & TML       \\
           & 0.070                             & 2.2               & C\&L 2003 &            & 0.18             & 0.86              & Z         \\
           & 0.21                              & 0.73              & This Work &            & 1.5              & 0.10              & S\&W      \\
B0834+06   & 0.122                             & 1.25              & TML       &            & 0.265            & 0.576             & C\&L 2003 \\
           & 0.454                             & 0.337             & BGR       &            & 0.55             & 0.28              & This Work \\
           & 0.21                              & 0.73              & Z         & B2016+28   & 0.0154           & 9.91              & KLL       \\
           & 0.60                              & 0.25              & S\&W      &            & 0.0287           & 5.32              & TML       \\
           & 0.197                             & 0.776             & C\&L 2003 &            & 0.206            & 0.742             & BGR       \\
           & 0.38                              & 0.40              & This Work &            & 0.18             & 0.86              & Z         \\
B0919+06   & 0.0228                            & 6.70              & KLL       &            & 0.11             & 1.35              & S\&W      \\
           & 0.00930                           & 16.4              & TML       &            & 0.046            & 3.3               & C\&L 2003 \\
           & 0.256                             & 0.597             & BGR       &            & 0.13             & 1.2               & This Work \\
           & 0.21                              & 0.73              & Z         & B2020+28   & 0.074            & 2.07              & KLL       \\
           & 0.19                              & 0.81              & S\&W      &            & 0.0614           & 2.49              & TML       \\
           & 0.015                             & 10                & C\&L 2003 &            & 0.270            & 0.566             & BGR       \\
           & 0.12                              & 1.2               & This Work &            & 0.23             & 0.67              & S\&W      \\
B0950+08   & 1.04                              & 0.147             & KLL       &            & 0.0986           & 1.55              & C\&L 2003 \\
           & 77.29                             & 0.001977          & TML       &            & 0.25             & 0.60              & This Work \\
           & \textgreater{}\textgreater{}9.000 & 0.01698           & BGR       & B2110+27   & 0.0181           & 8.43              & KLL       \\
           & 13                                & 0.012             & Z         &            & 0.0406           & 3.76              & TML       \\
           & 25                                & 0.0060            & S\&S      &            & 0.24             & 0.64              & Z         \\
           & 20                                & 0.0076            & This Work &            & 0.065            & 2.3               & C\&L 2003 \\
B1237+25   & 0.208                             & 0.735             & TML       &            & 0.062            & 2.5               & This Work \\
           & 1.828                             & 0.08358           & BGR       & B2315+21   & 0.00973          & 15.7              & KLL       \\
B2315+21   & 0.0773                            & 1.98              & TML       & J1612+2008 & 0.25             & 0.61              & This Work \\
           & 0.17                              & 0.88              & Z         & J1652+2651 & 0.0045           & 34                & KLL       \\
           & 0.124                             & 1.23              & C\&L 2003 &            & \textless{}0.084 & \textgreater{}1.8 & This Work \\
           & 0.17                              & 0.9               & This Work & J1758+3030 & 0.0022           & 69                & KLL       \\
J0051+0423 & 0.26                              & 0.59              & Z         &            & \textless{}0.084 & \textgreater{}1.8 & This Work \\
           & 0.37                              & 0.42              & This Work & J2215+1538 & \textless{}0.084 & \textgreater{}1.8 & This Work \\
J0137+1654 & 0.023                             & 6.6               & This Work & J2227+30   & 0.21             & 0.74              & This Work \\
J1313+0931 & 0.31                              & 0.49              & This Work & J2253+1516 & 0.030            & 5.0               & This Work \\
\enddata
\tablecomments{The reference keys are as follows:
\begin{itemize}
    \item KLL = \citet{kuzmin2007measurements}: KLL observed at 111~MHz and fit to the scattering tail with a truncated exponential. 
    \item TML = \citet{taylor1993catalog}: TML measured a scintillation bandwidth at 1~GHz using a correlation method. In actuality, the $\tau_s$ value in this paper was \textit{measured} as a scintillation bandwidth, but only \textit{reported} as a scattering time. Here we ``de-convert'' the reported value back to the original scintillation bandwidth before rescaling.
    \item DLK = \citet{daszuta2013scintillation}: DLK observed at 1.7~GHz and fit a 2D Gaussian to the ACF to determine scintillation bandwith and timescale.
    \item BGR = \citet{bhat1998bubble} observed at 327~MHz and fit a 2D Gaussian to the ACF to determine scintillation bandwith and timescale.
    \item Z = \citet{zakharenko2013detection} observed at 10--30~MHz and fit to the scattering tail with a boxcar-convolved exponential. 
    \item S\&W = \citet{smith1985frequency}. S\&W observed at 408 MHz and fit a 2D Gaussian to the ACF to determine scintillation bandwith and timescale.
    \item S\&S = \citet{Smirnova_2008}. S\&S measured a scintillation bandwidth at four frequencies (41, 62, 89, and 112 MHz) using correlation methods. The value we report here is the average.
    \item C\&L = \citet{cordes2003ne2001}. C\&L compiled scintillation bandwidths from various previous studies; here, all measurements are scintillation bandwidths originally measured in \citet{cordes1986space} at 1~GHz using the half-width-at-half maximum of the intensity correlation function of the dynamic spectrum. 
\end{itemize}}
\label{tab:literature}
\end{deluxetable}

 \label{tab:literature}

In Table \ref{tab:literature}, rows with ``This Work'' have been averaged from all of this work's observations \edit1{of each pulsar} using the result from the Gaussian fit (as justified in Section \ref{ssec:population}). Errors are not included in the values from Table \ref{tab:literature}, as different works have incompatible standards of presenting and calculating errors.

Finally, to evaluate trends in model over- or under-estimation for each pulsar's scintillation bandwidth, we quantified the amount of difference between the prediction and the measurement using a simple \ac{DF} calculated with:

\begin{equation}
    \rm{Difference\ Factor} = \frac{\Delta\nu_{D, measured} - \Delta\nu_{D, predicted}}{\Delta\nu_{D, predicted}}
\end{equation}

Note that we evaluate consistency in the following sections using 1-sigma errors. We comment upon a few interesting sub-populations and measurements in the subsections below.

\subsection{Pulsars with no literature values}
\label{ssec:no_literature}

We measured the scintillation bandwidths for the following pulsars for the first time, as there were no reported measurements in the literature in the form of either scintillation bandwidths or scattering timescales. We obtained a single measurement each for J2227+3038, J2253+1516, J0137+1654, J1313+0931, and J1612+2008, all of which were larger than the model predictions. In our dataset, we obtained two measurements of J2215+1538 --- one upper limit consistent with the narrow estimate from both models, and the other measured value larger than both models' predictions. 

\subsection{Pulsars with negative difference factor}
\label{ssec:negative_df}

Most pulsars that we measured were either consistent with or had larger scintillation bandwidth values than the model predictions, as will be discussed in Section \ref{ssec:population}. The following pulsars, in contrast, have measured scintillation bandwidth values which are \textit{smaller} than some of the model predictions.

We measured one observation for the scintillation bandwidth for B0950+08. The model predictions were quite different for this pulsar --- 101~MHz for NE2001 versus 4~MHz for YMW16 --- and the measured value fell between these bounds. We find that our measurement is inconsistent with that of previous literature \citep[e.g., ][]{taylor1993catalog, kuzmin2007measurements} but is consistent with others \citep[e.g., ][]{bhat1998bubble, zakharenko2013detection}, including, notably, a wide-bandwidth investigation of B0950+08 over four distinct frequency ranges \citep{Smirnova_2008}.

We measured four bandwidth values for B2110+27, which were generally consistent with each other and NE2001 and a few times larger than YMW16. The mean for both the Lorentzian and Gaussian measurements was 0.06$\pm$0.01~MHz. This is consistent with some prior reports in the literature \citep{taylor1993catalog, kuzmin2007measurements}, but not \citet{zakharenko2013detection}.

We made three measurements of B1929+10, two of which were consistent with each other within 1-sigma error. All three measurements were lower than the values predicted by NE2001 and YMW16. The measurement in \citet{taylor1993catalog} is consistent with our lowest measurement, while the measurement from \citet{smith1985frequency} falls within the range of our three measurements. The mean for the Lorentzian measurement was 1.3$\pm$0.6~MHz and for the Gaussian measurement was 1.2$\pm$0.6~MHz.

\subsection{Particularly narrow pulsars (upper limit only)}
\label{ssec:upper_limit}

Pulsars in this category have scintillation bandwidth measurements which are upper limits, indicating that the actual bandwidth is narrower than we can measure ($<$0.084 MHz, or five points across the ACF slice described in Section~\ref{ssec:fitting}). We set an upper limit on two observations for the scintillation bandwidth for B0301+19, which is consistent with both NE2001 and YMW16, as well as with \citet{kuzmin2007measurements} and \citet{taylor1993catalog} but not with \citet{smith1985frequency}, who found a larger value. We set an upper limit on one observation each for the scintillation bandwidths for J1652+2651 and J1758+3030 which were consistent with both NE2001 and YMW16, as well as \citet{kuzmin2007measurements} (the only measured value in the literature). J2215+1538 has one upper limit measurement and is discussed in full in \ref{ssec:no_literature}.

\subsection{Unexpected measurements based on sample selection}
\label{ssec:unexpected}

We did not expect to be able to measure bandwidths for these pulsars because their scintillation bandwidths were predicted to be too small for us to resolve at the observing frequencies. We were only able to find upper limits for J1758+3030 and one of two observations of J2215+1538, but scintillation was still visible in the dynamic spectra. For pulsars B0919+06, J0137+1654, J1652+2651, J2227+3038, J2253+1516, and the other observation of J2215+1538, we report measured bandwidths despite the fact that these pulsars were outside the error bars on the NE2001 prediction used to guide sample selection in Section~\ref{ssec:data_selection}.

\subsection{B0820+02}
\label{ssec:B0820+02}

We made two measurements of B0820+02, which are consistent with each other within 1-sigma error. Both measurements are larger than the model predictions, especially compared to YMW16. \citet{taylor1993catalog} was more consistent with the model predictions, while the lower limit from \citep{smith1985frequency} was an order-of-magnitude larger than the measurements in this work. It should be noted that B0820+02 is in a binary system with a white dwarf companion of mass 0.6~$M_{sun}$ \citep{koester2000white}; in the initial folding and assessment with \texttt{prepfold}, we observed no significant deviation from vertical in the time-phase plots, indicating that the effect of binarity on the data were minimal. Therefore, we used the simple, non-ephemeris solution when analyzing this pulsar. B0820+02 is also of interest due to its nulling and sub-pulse drifting behaviour \citep{Zhi2023B0820+20}.

\subsection{B0919+06}
\label{ssec:B0919+06}

According to model predictions, we should not have been able to resolve scintles for this pulsar at our observing frequency. However, we obtained three measurements of B0919+06, all an order-of-magnitude larger than predictions from both YMW16 and NE2001, two of which were consistent within error. These measurements were generally larger than \citet{kuzmin2007measurements} and \citet{taylor1993catalog}, but smaller than \citet{bhat1998bubble}, \citet{zakharenko2013detection}, and \citet{smith1985frequency}. The mean for the Lorentzian measurement was 0.13$\pm$0.01~MHz, and for the Gaussian measurement was 0.13$\pm$0.02 \edit1{MHz}.

\subsection{B2020+28}
\label{ssec:B2020+28}

We measured three scintillation bandwidths for B2020+28, two of which were consistent with each other, but not with NE2001 and YMW16 predictions. These measurements were consistent with \citet{smith1985frequency}, but not \citet{bhat1998bubble}, \citet{taylor1993catalog}, or \citet{kuzmin2007measurements}. The mean for the Lorentzian measurement was 0.27$\pm$0.01~\edit1{MHz} and for the Gaussian measurement was 0.25$\pm$0.02~MHz.

\subsection{Population Studies}
\label{ssec:population}

\begin{figure}
    \centering
    \includegraphics[width=\textwidth]{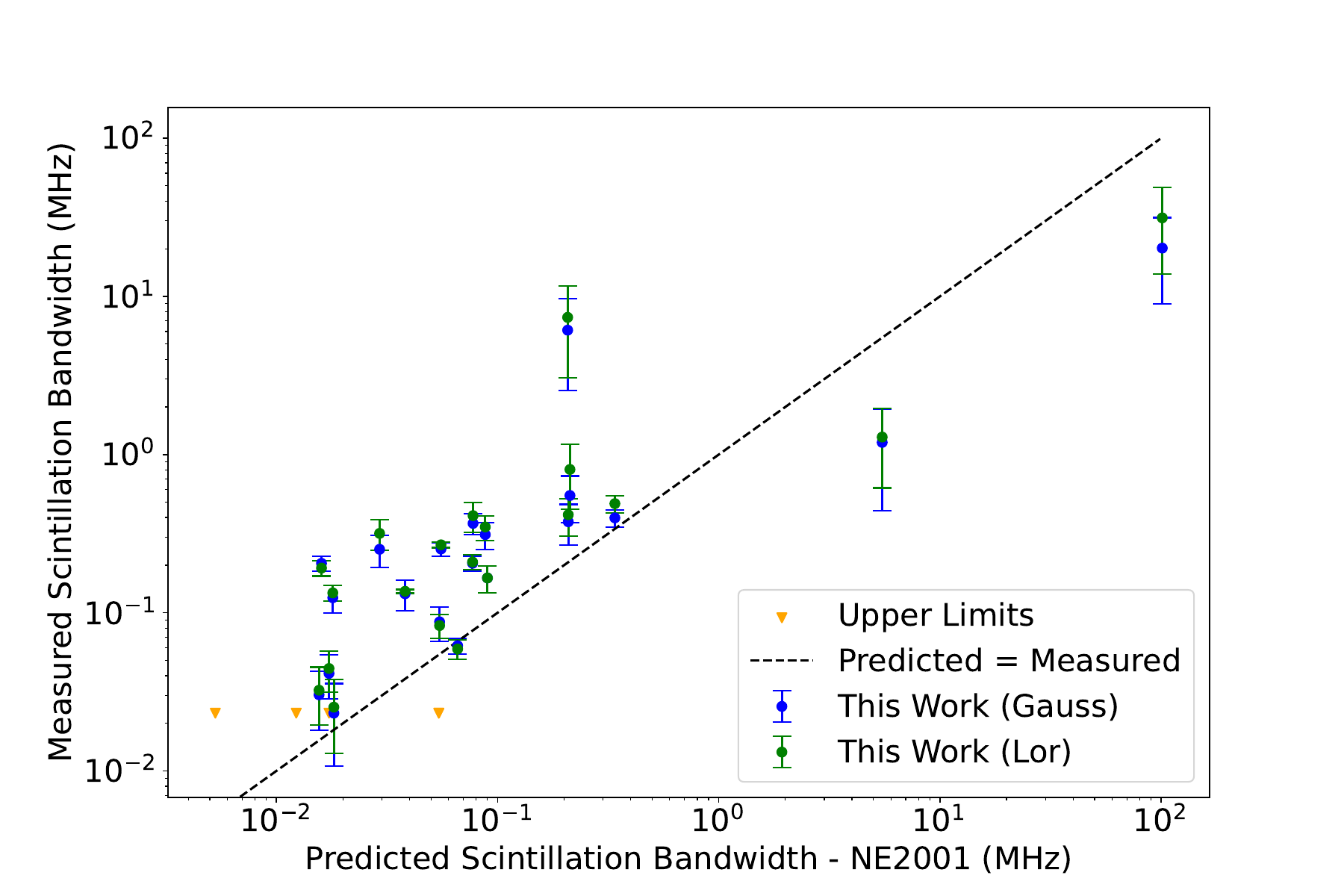}
    \caption{Predicted scintillation bandwidths from NE2001 plotted against the measured scintillation bandwidths from this study fitted with Gaussians (blue) and Lorentzians (green). Upper limits are shown with gold triangles. In general, the measurements were larger than their corresponding predictions. 1-sigma errors are displayed for the measured scintillation bandwidth, as calculated in Section~\ref{ssec:errors}.}
    \label{fig:predicted_vs_measured}
\end{figure}

\begin{figure}
    \centering
    \includegraphics[width=\textwidth]{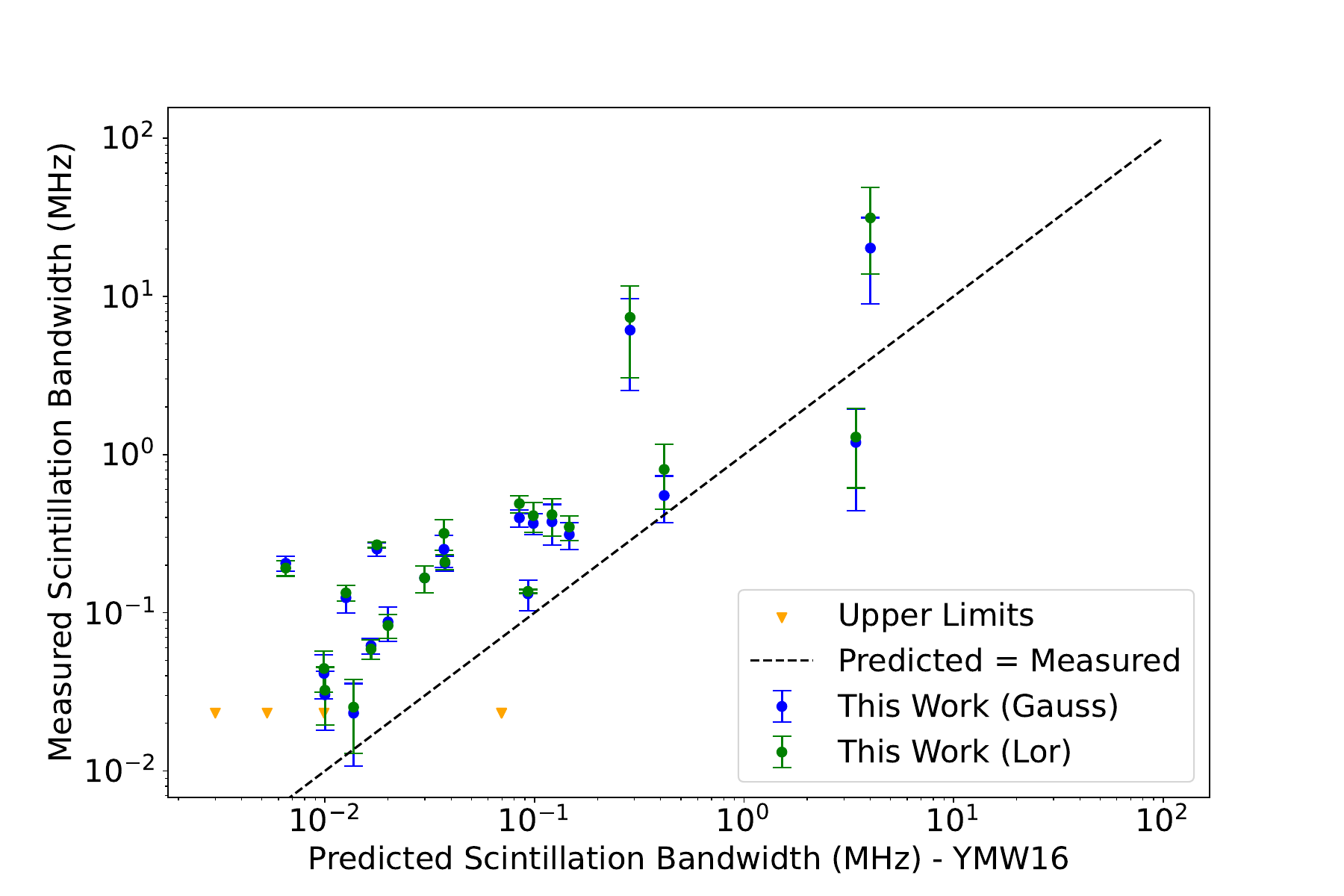}
    \caption{Predicted scintillation bandwidths using \citet{krishnakumar2015scatter} (YMW16) plotted against the measured scintillation bandwidths from this study fitted with Gaussians (blue) and Lorentzians (green). Upper limits are shown with gold triangles. Even more strongly than with NE2001, the measurements were larger than their corresponding predictions. 1-sigma errors are displayed for the measured scintillation bandwidth, as calculated in Section~\ref{ssec:errors}.}
\label{fig:predicted_vs_measured_ymw}
\end{figure}

Our measurements compare to the NE2001 predictions as shown in Figure \ref{fig:predicted_vs_measured}. In almost every case, the measurements that we report in Section \ref{sec:results} are larger than their corresponding NE2001 predictions. A few observations showed clear scintillation in the dynamic spectra, but did not produce a peak in the ACF wider than five points across the profile, implying that there was a scintillation bandwidth, but that it was too narrow to resolve with our frequency resolution. These five measurements are indicated in the gold triangles in Figure \ref{fig:predicted_vs_measured}. Note that these are reported with a generous 0.084~MHz upper limit in Table~\ref{tab:measurements}; some measured values in other pulsars are narrower than this upper limit due to the difference between the total width across the cropped slice and the extracted width parameter of the fitted function. We do the same analysis for YMW16 predictions in Figure \ref{fig:predicted_vs_measured_ymw}, and find an even stronger underprediction than with NE2001.

\edit1{At this point, it should be noted that the predictions themselves make inherent assumptions about the frequency scaling of scintillation bandwidths and their relation to scattering delays. In this work, both the NE2001 predictions and the YMW16 predictions were obtained by using a Kolmogorov scaling ($\alpha$ = 4.4) to shift the predictions from the default value of 1~GHz to 327~MHz. NE2001 did use $\alpha$ = 4.4 in the model creation, but YMW16 used the square-law scaling $\alpha$ = 4.0 instead. We also tried an $\alpha$ = 4.0 re-scaling of predicted bandwidths for both NE2001 and YMW16 and found hints that the scaling is shallower than 4.4. However, none of the conclusions in the rest of this section are affected by the change in scaling, and it is not consistent to \textit{post facto} adjust and compare the scalings without going back to the measurements and observing frequencies that each model is based on and rescaling.}

\begin{figure}
    \centering
    \includegraphics[width=0.5\textwidth]{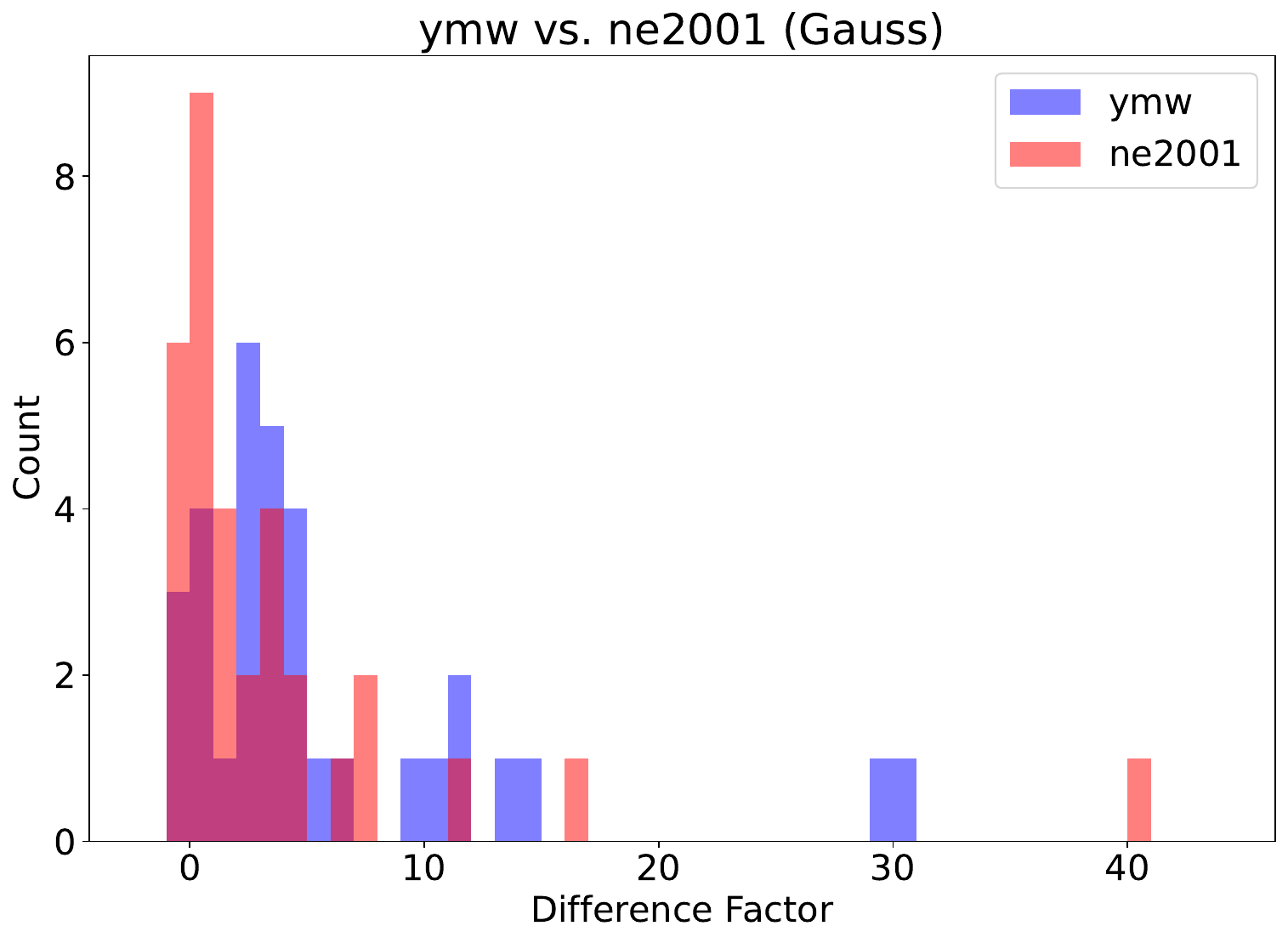}
    \caption{A histogram comparing the distribution of ``difference factor'' (the consistency of the model with the data) for YMW16 versus NE2001 predictions compared to data with Gaussian fits. Even with a small sample size, it is clear that the NE2001 distribution peaks closer to zero than the YMW16 distribution in the lefthand plot, indicating that NE2001 is a better fit to the data.}
    \label{fig:bestmodel}
\end{figure}

\begin{figure}
    \centering
    \includegraphics[width=0.5\textwidth]{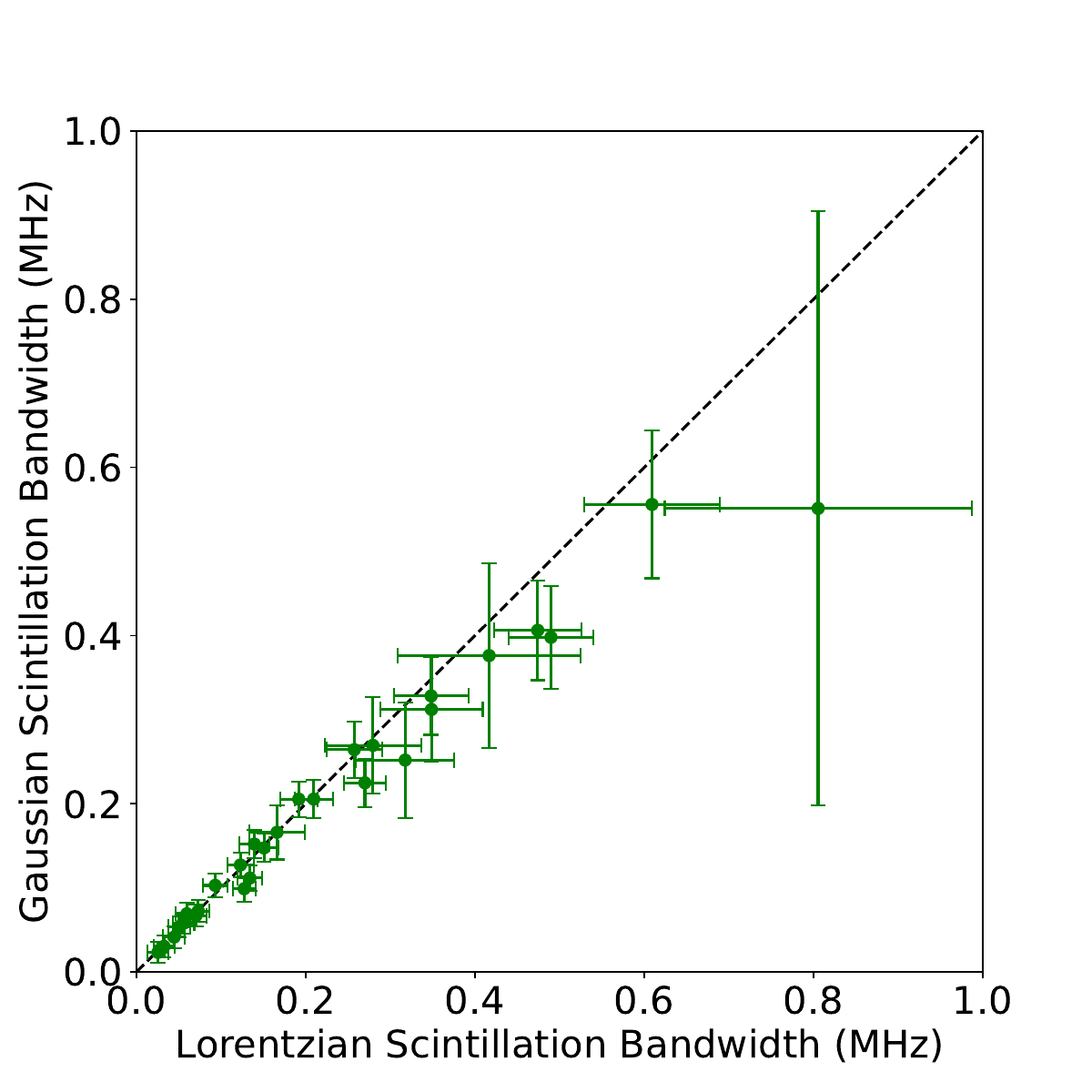}
    \caption{A scatter plot comparing the Lorentzian (x-axis) and Gaussian (y-axis) scintillation bandwidth measurements below 1~MHz. The Gaussian and Lorentzian measurements for each scintillation bandwidth are generally similar, but the Lorentzians start trending larger at larger scintillation bandwidths (this trend does continue into the four measurements above 1~GHz, omitted for plot readability). 1-sigma errors are displayed for each fit, as calculated in Section~\ref{ssec:errors}}
    \label{fig:bestfit}
\end{figure}

Given that the measurements consistently deviated from the predictions, we evaluated whether these deviations were systematically linked to any physically-meaningful parameter. Remember that the \ac{DF} is dependent upon which model $\nu_{D, \mathrm{predicted}}$ (NE2001 or YMW16) is being used, as well as which fit function $\nu_{D, \mathrm{measured}}$ is selected for the reported bandwidth. We found that the combination of model and fit function with the lowest median \ac{DF} was a Gaussian model fit combined with NE2001 predictions, with a median \ac{DF} of 1.59. A Lorentzian model fit with YMW16 predictions was the worst of the options, with a median \ac{DF} of 3.49, while a Gaussian model fit with YMW16 predictions was only slightly better, with a median \ac{DF} of 3.16. It seems clear that NE2001 is a better match to our data than YMW16 (see Figure \ref{fig:bestmodel}). The difference between the Gaussian and Lorentzian fits, assuming NE2001, is actually quite small; a Lorentzian model fit with NE2001 predictions produces a median \ac{DF} of 1.72. This is shown graphically in Figure \ref{fig:bestfit}, where the Gaussian and Lorentzian measurements are generally similar, except that the Lorentzian measurements trend larger than the Gaussian measurements at larger scintillation bandwidths. A slight preference for the Gaussian fit is expected; previous measurements and models, including NE2001, often made the Gaussian approximation. 

In addition, if we divide our sample into pulsars which were used for training NE2001 \citep[as in Table 3 of ][]{cordes2003ne2001}, and pulsars which were not --- conveniently divided along the lines of ``B'' and ``J'' labelled pulsars respectively in our Table \ref{tab:measurements} --- we find that the pulsars which were used for training NE2001 have a median \ac{DF} of 0.90 (NE2001, Gaussian), as opposed to the pulsars which were not used for training NE2001, which have a median \ac{DF} of 2.89 (NE2001, Gaussian). We thus confirm that pulsars which were used for training NE2001 generally conform more closely to the NE2001 model predictions.

\begin{figure}
    \centering
    \includegraphics[width=\textwidth]{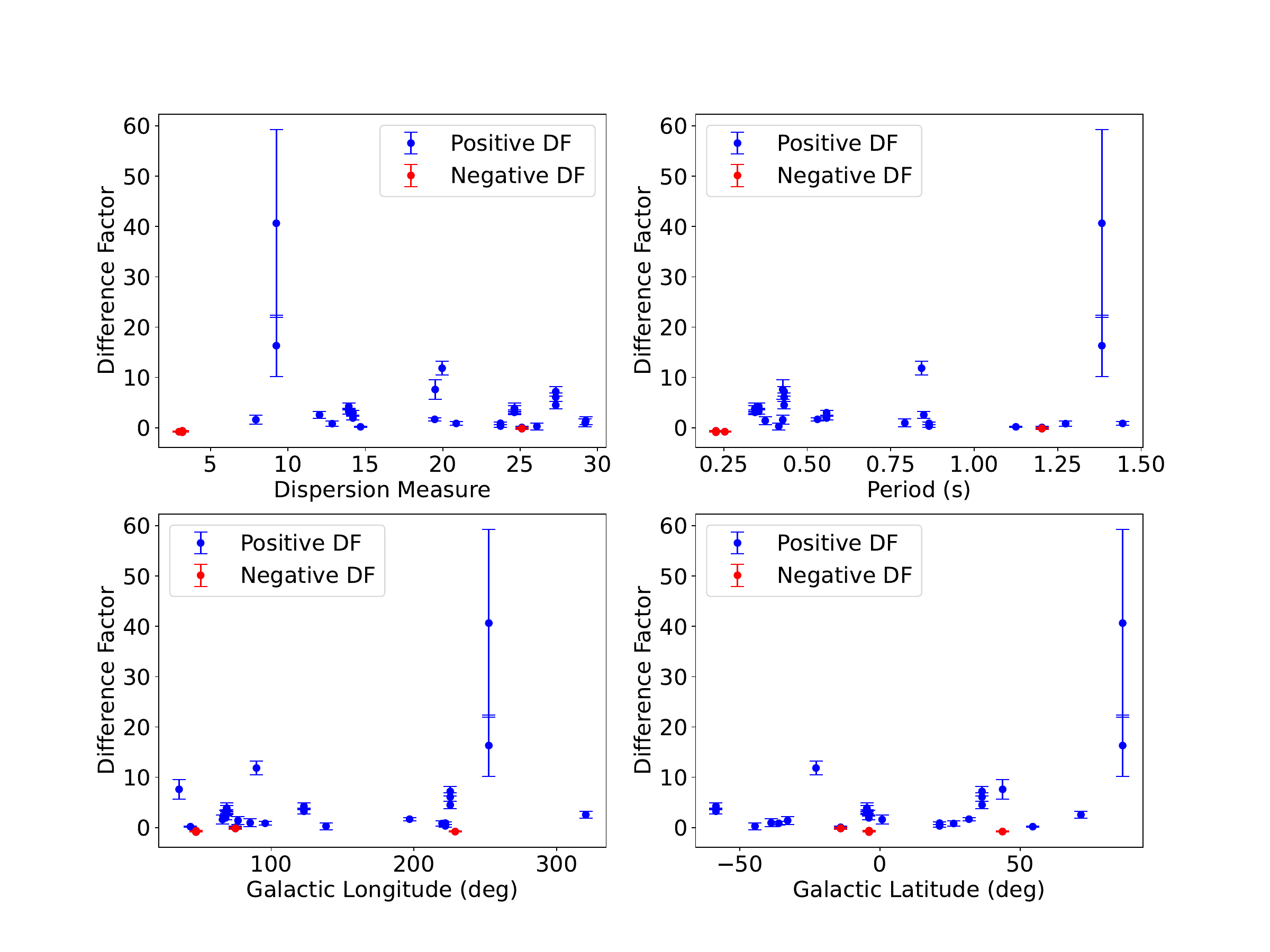}
    \caption{Correlations between difference factor (DF; Gaussian fit vs. NE2001) and DM (top left), period (top right), galactic longitude (bottom left) and galactic latitude (bottom right). While no universal trends are visible, the presence of measurements with negative difference factor at low DMs and periods, and the high difference factor on the pulsar furthest from the galactic plane, could point to differing accuracy of NE2001 in certain directions and distances. 1-sigma errors are displayed for the difference factor, calculated using only measurement errors from Section~\ref{ssec:errors} (not model errors). The two points with large DF and errorbars near 250$^{\circ}$ galactic longitude belong to B1237+25, whose measured scintillation bandwidths were significantly larger than the prediction, compared to the rest of the sample. Large measured bandwidths are associated with larger absolute errors, which are exaggerated in this case by the small predicted bandwidth.}
    \label{fig:correlations}
\end{figure}

Given that the combination of NE2001 and Gaussian has the lowest average \ac{DF}, we plotted the \ac{DF} for a Gaussian fit versus NE2001 against the \ac{DM}, spin period $P$, galactic latitude, and galactic longitude for each pulsar as shown in Figure \ref{fig:correlations}. We also visually inspected all four possible combinations of model and fit function, determining that there is no difference in correlations between any of the four. 

We do not see any general trends or linear correlations between \ac{DF} and any of the variables, however, some interesting conclusions can still be drawn from the rare cases where the prediction exceeded the measurement (shown in red), or where the \ac{DF} was highest i.e., where the measurement far exceeded the prediction.

From the \ac{DM} and period subplots, we note that the measurements with negative \ac{DF} are associated with pulsars that are nearby and have short spin periods. However, only three pulsars show measurements with negative \ac{DF}, and only two of those, B1929+10 and B0950+08 are driving this conclusion (albeit two pulsars that are in very different parts of the sky). Finally, we note the pulsar with the largest two measurements of \ac{DF}, B1237+25, is located extremely far from the galactic plane at a galactic latitude of 86.54$^\circ$. Given that there are far more galactic radio sources within the galactic plane, it is reasonable that electron density models may be less accurate far from the plane.

\begin{figure}
    \centering
    \includegraphics[width=0.7\textwidth]{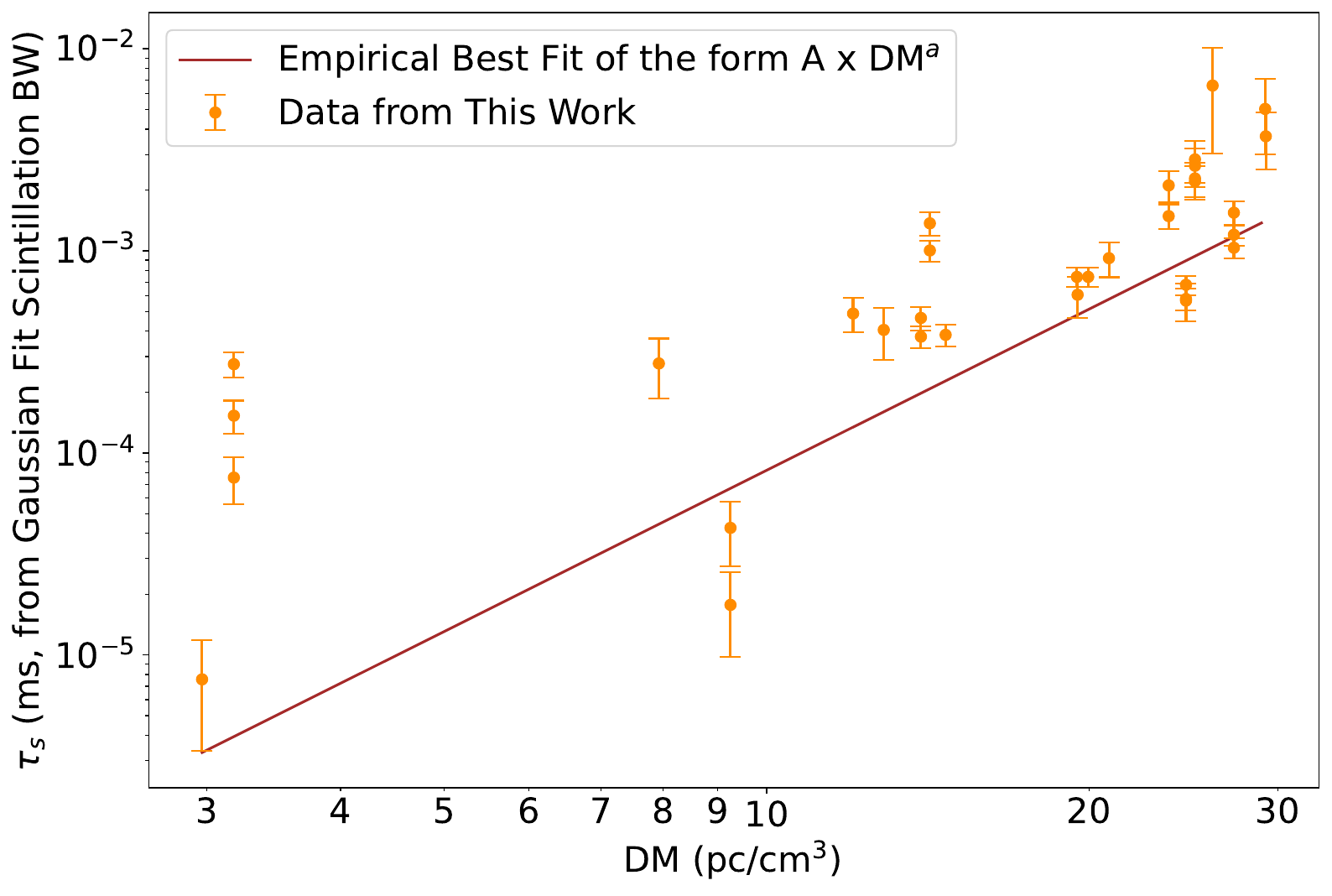}
    \caption{Dispersion measure versus scattering time in ms, shown on logarithmic axes. \edit1{The orange points indicate the DM and $\tau_{s}$ values from our measurements, while the brown line shows the best fit of the form $A \times \mathrm{DM}^a$. We find that the \ac{DM} scaling method somewhat matches our dataset, with best fit parameters of A = $2.1 \times 10^{-1}$ and a = 2.6, but there is still significant scatter.} We display 1-sigma errors for the converted scattering time, as calculated in Section~\ref{ssec:errors}. \edit2{Note that the sizes of the errorbars span four orders-of-magnitude, and points with smaller errorbars are significantly upweighted in the fit; this can be difficult to gauge visually given that points with different magnitudes of errorbars can look deceptively similar in log-log plotting.}}
    \label{fig:dm_taus.pdf}
\end{figure}

\edit1{Finally, we investigate the possibility of a \ac{DM}--$\tau_{s}$ relation in our data by converting from scintillation bandwidth as in Equation \ref{eq:scattering_to_bandwidth}, with $C=0.96$ (Figure \ref{fig:dm_taus.pdf}). Both \citet{krishnakumar2015scatter} and \citet{cordes2016radio} empirically fit their data with a relation of the form $\tau_{s} \mathrm{(ms)}$ = $A \times \mathrm{DM}^a (1 + B \times \mathrm{DM}^b)$ \citep[as in][]{ramachandran1997kinematics} where the parenthetical component of the equation (i.e., the steeper positive slope in a log-log plot) becomes dominant for pulsars with \ac{DM}s greater than a few dozen. In this work, where the largest \ac{DM} is 29.24~pc~cm$^{-3}$, we opt to fit a simpler equation of the form $\tau_{s} \mathrm{(ms)}$ = $A \times \mathrm{DM}^a$ using the \texttt{curve\_fit} function from \texttt{scipy.optimize} with errors incorporated.} \edit2{We find $A = 1.83 \times 10^{-7}$ and $a = 2.65$. \citet{cordes2016radio} found that $A = 2.98 \times 10^{-7}$, and $a = 1.4$, while \citet{krishnakumar2015scatter} found that $A = 3.6 \times 10^{-6}$ and $a = 2.2$. Our $A$ value is thus smaller than both previous references, although more consistent with \citet{cordes2016radio}, and our $a$ value is larger than both previous references, though more consistent with \citet{krishnakumar2015scatter}.} \edit1{Figure~\ref{fig:dm_taus.pdf} indicates that the DM scaling method can be applied to this dataset with some success, but the large residuals indicate that additional effects are likely present.}

\section{Discussion and Conclusion}
\label{sec:discussion_and_conclusion}

In this work, we used archival drift scan data from the Arecibo telescope's AO327 survey to measure 38 scintillation bandwidths ($\Delta \nu_D$) from 23 pulsars, six of which have no existing $\Delta \nu_D$ values in the literature. We fit Lorentzian and Gaussian models to the \acp{2D ACF} of the dynamic spectra from these observations, as reported in Section \ref{sec:analysis}, and compared them to the predictions from the NE2001 and YMW16 models of galactic electron density. Generally, in Section \ref{sec:results}, we found that our measurements were larger than those predicted by both electron density models. We required at least five points across the profile for the Gaussian and Lorentzian fits, though underresolution of the profile could still play a role in the general underestimation of the models. We found that NE2001 provided smaller difference factors (measuring the difference between the data and the model prediction) than YMW16. We also found that the Gaussian fits were, on average, closer to the model predictions than the Lorentzian fits. While there should be a Fourier relationship between the Lorentzian and the assumed exponential response function of the \ac{ISM} \citep{turner2024simultaneous}, Gaussians have traditionally been used for fitting throughout the literature, including for NE2001's training pulsars. The significant presence of NE2001 training pulsars in this dataset thus potentially explains the success of the Gaussian fit.

Using Gaussian fits, we found that the scintillation bandwidths measured for three  nearby millisecond pulsars were smaller than NE2001 predictions, breaking the trend set by the other 20 pulsars (see Section \ref{ssec:negative_df}). In addition, four pulsars showed scintillation that was too narrow to measure (reported as an upper limit; see Section \ref{ssec:upper_limit}). We see no general trend with difference factor (Gaussian fit, NE2001) versus DM, period, or galactic coordinates, but do see some indication that NE2001 is less accurate far from the galactic plane and at low distances from Earth (Section \ref{ssec:population}). 

\edit1{At a high level, there is a question of whether direction-agnostic \ac{DM} scaling approaches \citep[e.g., ][]{krishnakumar2015scatter} or the modeling of large-scale distribution of interstellar scattering \citep[e.g., ][]{cordes2002NE2001} more successfully predict scintillation bandwidths. Here, despite the challenges of modeling electron density geometrically, NE2001 produces a better fit to the data i.e., smaller difference factors. These measurements could be used to inform the distribution of interstellar scattering and turbulence properties in the next generation of geometric models.}

In Section \ref{sec:results}, specifically the literature review in Table~\ref{tab:literature}, we also find that previous $\Delta \nu_D$ values for these pulsars in the literature are generally in agreement with each other and with our measurements in order-of-magnitude, but often disagree by factors of a few, barring some egregious outliers. Some of the disagreement could be due to the assumption of a Kolmogorov scaling of $\alpha$=4.4, which was used to make various sources compatible. More of it is likely due to true changes in the scintillation bandwidth over the years and decades between references, as scintillation bandwidths can change dramatically even over months-long timescales \citep[e.g., ][]{Hemberger2008variability}. Even so, for pulsars that had three or more measurements in our sample, we do not see any indication that observations taken closer together in time have better agreement in their scintillation bandwidth measurements than observations taken far apart. To better investigate this phenomenon, future work could more thoroughly account for \ac{DM} changes over time, as discussed in \citet{shapiro2021study}.

This work covers about 97\% of the PUPPI data from the AO327 project, which were searched for a subset of 40\% of the pulsars in the \citet{deneva2024ao327} sample. With the pipeline described in Section \ref{sec:analysis} fully implemented, this pilot survey can be expanded to the full set of AO327 pulsar detections, including the earlier data taken with the Mock Spectrometer, providing a uniform sample against which to compare future $\Delta \nu_D$ measurements in the literature. This is increasingly relevant due to the connections between better pulsar timing and the detection and characterization of low-frequency gravitational waves.

\begin{acronym}
    \acro{ISM}{interstellar medium}
    \acro{DISS}{diffractive interstellar scintillation}
    \acro{RISS}{refractive interstellar scintillation}
    \acro{FRB}{fast radio burst}
    \acro{PUPPI}{Puerto Rico Ultimate Pulsar Processing Instrument}
    \acro{RFI}{radio frequency interference}
    \acro{FWHM}{full-width at half-maximum}
    \acro{RRAT}{rotating radio transient}
    \acro{PRESTO}{the PulsaR Exploration and Search TOolkit}
    \acro{SNR}{signal-to-noise ratio}
    \acro{DM}{dispersion measure}
    \acro{ACF}{autocorrelation function}
    \acro{FSE}{finite scintle error}
    \acro{DF}{difference factor}
    \acro{2D ACF}{2D autocorrelation function}
    \acro{HWHM}{half-width at half-maximum}
\end{acronym}

\acknowledgments{
S.Z.S. acknowledges that this material is based upon work supported by the National Science Foundation MPS-Ascend Postdoctoral Research Fellowship under Grant No. 2138147. This work was performed as part of the Penn State Pulsar Search Collaboratory, which is an offshoot of the Pulsar Search Collaboratory project funded under NSF award No. 151651. This project used archival data from the AO327 project, funded under NSF award No. 2009425, and utilized resources from the NANOGrav Physics Frontiers Center, funded under NSF award No. 2020265. M.T.L. acknowledges support received from NSF AAG award number 2009468, and NSF Physics Frontiers Center award number 2020265, which supports the NANOGrav project. M.A.M. is supported by NSF Physics Frontiers Center award number 2020265 and NSF award number 2009425. Figure 1 was created using the SankeyMATIC software (\url{https://github.com/nowthis/sankeymatic})}

\bibliographystyle{aasjournal}
\bibliography{references}

\end{document}